\RequirePackage{lineno}
\documentclass[prd,twocolumn,showpacs,amsmath,amssymb]{revtex4}
\usepackage{graphicx}
\usepackage{dcolumn}
\usepackage{bm}
\usepackage{epsfig}
\usepackage{overpic}
\usepackage{verbatim}
\newsavebox{\tablebox}
\usepackage[colorlinks,linkcolor=blue,anchorcolor=red,citecolor=blue]{hyperref}

\usepackage{rotating}

\newcommand {\degree}   {${}^{\circ}$}
\def\ra{\ensuremath{\rightarrow}}
\newcommand{\br}[1]{\mathcal{B}_{#1}}
\newcommand{\ee}{e^+e^-}

\def\GeV{\ifmmode {\mathrm{\ Ge\kern -0.1em V}}\else
                   \textrm{Ge\kern -0.1em V}\fi}%
\def\MeV{\ifmmode {\mathrm{\ Me\kern -0.1em V}}\else
                   \textrm{Me\kern -0.1em V}\fi}%
\let\gev=\GeV
\let\mev=\MeV

\def\GeVc{\ifmmode {\mathrm{\ Ge\kern -0.1em V}/c}\else
                   {\textrm{Ge\kern -0.1em V}/$c$}\fi}%
\def\MeVc{\ifmmode {\mathrm{\ Me\kern -0.1em V}/c}\else
                   {\textrm{Me\kern -0.1em V}/$c$}\fi}%

\let\mevc=\MeVc

\def\GeVcc{\ifmmode {\mathrm{\ Ge\kern -0.1em V}/c^2}\else
                   {\textrm{Ge\kern -0.1em V}/$c^2$}\fi}%
\def\MeVcc{\ifmmode {\mathrm{\ Me\kern -0.1em V}/c^2}\else
                   {\textrm{Me\kern -0.1em V}/$c^2$}\fi}%
\let\gevcc=\GeVcc

\def\cm{\ifmmode  {\mathrm{\ cm}}\else
                   \textrm{~cm}\fi}%
\def\ifb{\mbox{fb$^{-1}$}}
\def\ipb{\mbox{pb$^{-1}$}}
\def\gam{\gamma}
\def\piz{\pi^0}
\def\pip{\pi^+}
\def\pim{\pi^-}
\def\pipm{\pi^{\pm}}
\def\pimp{\pi^{\mp}}
\def\kp{K^+}%
\def\km{K^-}%
\def\kpm{K^{\pm}}%
\def\ksz{K^0_{S}}
\def\klz{K^0_{L}}
\def\kz{K^0}
\def\kzbar{\bar{K}^0}
\newcommand{\etap}{\eta^{\prime}}
\newcommand{\Dz}{D^{0}}

\newcommand{\Dst}{D^*}

\newcommand{\Ds}{D_{s}}
\newcommand{\Dsp}{D_{s}^{+}}
\newcommand{\Dsm}{D_{s}^{-}}
\newcommand{\Dspm}{D_{s}^{\pm}}

\newcommand{\Dsst}{D_{s}^{*}}
\newcommand{\Dsstp}{D_{s}^{*+}}
\newcommand{\Dsstm}{D_{s}^{*-}}

\newcommand{\DspDsm}{D_{s}^{+}D_{s}^{-}}

\newcommand{\pkkpi}{K^{+}K^{-}\pi^{+}}
\newcommand{\mkkpi}{K^{+}K^{-}\pi^{-}}

\newcommand{\mkkpipiz}{K^{+}K^{-}\pi^{-}\pi^{0}}
\newcommand{\pksk}{K^0_{S}K^{+}}
\newcommand{\mksk}{K^0_{S}K^{-}}

\newcommand{\pklk}{K^0_{L}K^{+}}

\newcommand{\mpipipi}{\pi^{+}\pi^{-}\pi^{-}}

\newcommand{\mksktpi}{K^0_{S}K^{+}\pi^{-}\pi^{-}}

\newcommand{\mkskpipi}{K^0_{S}K^{-}\pi^{+}\pi^{-}}

\newcommand{\mkskpiz}{K^0_{S}K^{-}\pi^0}

\newcommand{\mkskspi}{K^0_{S}K^0_{S}\pi^{-}}

\newcommand{\mkpipi}{K^{-}\pi^{+}\pi^{-}}

\newcommand{\mpietathpi}{\pi^{-}\eta_{\pi^{+}\pi^{-}\pi^0}}

\newcommand{\mpietaprhog}{\pi^{-}\eta'_{\gamma\rho^0}}

\newcommand {\PR}     {Phys.{} Rev.{} }
\newcommand {\PRL}    {Phys.{} Rev.{} Lett.{} }
\newcommand {\PRep}   {Phys.{} Rep.{} }
\newcommand {\JHEP}   {J.{} High{} Energy{} Physics{}}
\newcommand {\NIM}    {Nucl.{} Instrum.{} Meth.{} }
\newcommand {\PL}     {Phys.{} Lett.{} }

\begin{document}
\normalsize
\parskip=5pt plus 1pt minus 1pt

\title{\boldmath Study of the Decays $\Dsp\ra \ksz\kp$ and $\klz\kp$ }
\author{\small
M.~Ablikim$^{1}$, M.~N.~Achasov$^{10,d}$, P.~Adlarson$^{59}$, S. ~Ahmed$^{15}$, M.~Albrecht$^{4}$, M.~Alekseev$^{58A,58C}$, A.~Amoroso$^{58A,58C}$, F.~F.~An$^{1}$, Q.~An$^{55,43}$, Y.~Bai$^{42}$
% ...
\\
\vspace{0.2cm}
(BESIII Collaboration)\\
\vspace{0.2cm} {\it
$^{1}$ Institute of High Energy Physics, Beijing 100049, People's Republic of China\\
% ...
$^{4}$ Bochum Ruhr-University, D-44780 Bochum, Germany\\
% ...
$^{10}$ G.I. Budker Institute of Nuclear Physics SB RAS (BINP), Novosibirsk 630090, Russia\\
% ...
$^{15}$ Helmholtz Institute Mainz, Johann-Joachim-Becher-Weg 45, D-55099 Mainz, Germany\\
% ...
$^{42}$ Southeast University, Nanjing 211100, People's Republic of China\\
$^{43}$ State Key Laboratory of Particle Detection and Electronics, Beijing 100049, Hefei 230026, People's Republic of China\\
% ...
$^{55}$ University of Science and Technology of China, Hefei 230026, People's Republic of China\\
% ...
$^{58}$ (A)University of Turin, I-10125, Turin, Italy; (B)University of Eastern Piedmont, I-15121, Alessandria, Italy; (C)INFN, I-10125, Turin, Italy\\
$^{59}$ Uppsala University, Box 516, SE-75120 Uppsala, Sweden\\
% ...
$^{d}$ Also at the Novosibirsk State University, Novosibirsk, 630090, Russia\\
% ...
}
}


\begin{abstract}

Using an $\ee$ annihilation data sample corresponding to an integrated
luminosity of 3.19~$\ifb$ and collected at a center-of-mass energy $\sqrt{s} = 4.178~\gev$ with the BESIII detector, we  measure the absolute branching fractions $\br{}(\Dsp\ra\pksk) = (1.425\pm0.038_{\rm stat.}\pm0.031_{\rm syst.})\%$ and $\br{}(\Dsp\ra\klz \kp) =(1.485\pm0.039_{\rm stat.}\pm0.046_{\rm syst.})\%$.
The branching fraction of $\Dsp\ra\pksk$ is compatible with the world average and that of $\Dsp\ra\klz\kp$ is measured for the first time. We present the first measurement of the $\ksz$-$\klz$ asymmetry in the decays $\Dsp\ra \kz_{S,L}\kp$, and $R(\Dsp\ra \kz_{S,L}\kp)=\frac{\br{}(\Dsp\ra \pksk) -\br{}(\Dsp \ra \klz \kp )}{\br{}(\Dsp\ra \pksk) +\br{}(\Dsp \ra \klz \kp )}= (-2.1\pm1.9_{\rm stat.}\pm1.6_{\rm syst.})\%$. In addition, we measure the direct $CP$ asymmetries $A_{\rm CP}(\Dspm\ra\ksz\kpm) =  (0.6\pm2.8_{\rm stat.}\pm0.6_{\rm syst.})\%$ and $A_{\rm CP}(\Dspm\ra\klz\kpm) = (-1.1\pm2.6_{\rm stat.}\pm0.6_{\rm syst.})\%$.
\end{abstract}


\maketitle

\section{INTRODUCTION}

Two-body hadronic decays of charmed mesons, $D\ra P_1 P_2$ (where $P_{1,2}$ denotes a pseudoscalar meson), serve as an ideal environment to improve our understanding of the weak and strong interactions because of their relatively simple topology~\cite{Chau:1983du,Chau:1986du}.
Charmed-meson decays into hadronic final states that contain a neutral kaon are particularly attractive. Bigi and Yamamoto~\cite{Bigi:1994aw} first pointed out that the interference of the decay amplitudes of the Cabibbo-favored (CF) transition $D\ra  \kzbar  \pi $ and the doubly-Cabibbo-suppressed (DCS) transition  $D\ra  \kz  \pi $ can result in a measurable  $\ksz$-$\klz$ asymmetry
\begin{eqnarray}
R(D\ra  \kz_{S,L} \pi ) = \frac{\br{}(D\ra \ksz \pi  )-\br{}(D \ra \klz \pi )}{\br{}(D\ra \ksz \pi  )+\br{}(D\ra \klz \pi )}.
\label{eq:RD}
\end{eqnarray}
A similar asymmetry can be defined in $\Dsp$ decays by replacing $\pi$ with $K$.  Additionally,  as pointed out in Ref.~\cite{Yu:2017oky}, the interference between CF and DCS amplitudes can also lead to a new $CP$ violation effect, which is estimated to be of an order of $10^{-3}$.
The measurement of $\ksz$-$\klz$ asymmetries and $CP$ asymmetries in charmed-meson decays with neutral kaon provides insight into the DCS process, as well as information to explore $\Dz$-$\bar{D}^0$ mixing, $CP$ violation and SU(3) flavor-symmetry breaking effects in the charm sector~\cite{Xing:1996pn,Gershon:2015xra}.

On the theory side, different phenomenological models give predictions for the $\ksz$-$\klz$ asymmetries: the topological-diagrammatic approach~\cite{Chau:1986du} under the SU(3) flavor symmetry (DIAG) or incorporating the SU(3) breaking effects ($\rm SU(3)_{FB}$)~\cite{Cheng:2010ry,Bhattacharya:2009ps,Muller:2015lua}, the QCD factorization approach (QCDF)~\cite{Gao:2014ena}, and the factorization-assisted topological-amplitude (FAT)~\cite{Wang:2017ksn}.
The predicted $\ksz$-$\klz$ asymmetries in charmed-meson decays from these different approaches, as well as the measured values reported by the CLEO Collaboration~\cite{He:2007aj} are summarized in Table~\ref{tab:theorypre}. Considering the large range of values predicted for the $\ksz$-$\klz$ asymmetries, their measurements provide a crucial constraint upon models of the dynamics of charmed meson decays.

\begin{table*}
\caption{Predictions for $\ksz$-$\klz$ asymmetries in charmed-meson decays from different phenomenological models and the CLEO measurements.}
\label{tab:theorypre}
\begin{center}
\begin{tabular}{lccccccc}
\hline
\hline
 &DIAG~\cite{Bhattacharya:2009ps}            &      DIAG~\cite{Cheng:2010ry}           & QCDF~\cite{Gao:2014ena}   &$\rm SU(3)_{FB}$~\cite{Muller:2015lua}  &FAT~\cite{Wang:2017ksn} & CLEO~\cite{He:2007aj}\\
\hline
$R(D^0\ra \kz_{S,L}\piz)(\%)$ & 10.7					&10.7					&10.6			&$9^{+4}_{-2}$  			&$11.3\pm0.1$		&$10.8\pm2.5_{\rm stat.}\pm2.4_{\rm syst.}$\\
$R(D^+\ra \kz_{S,L}\pip)(\%)$ &$-0.5\pm1.3$			&$-1.9\pm1.6$				&$-1.0\pm2.6$		&-						&$2.5\pm0.8$		&$2.2\pm1.6_{\rm stat.}\pm1.8_{\rm syst.}$\\
$R(D_s^+\ra \kz_{S,L}\kp)(\%)$ &$-0.22\pm0.87$                       	&$-0.8\pm0.7$                       	&$-0.8\pm0.7$ 	&$11^{+4}_{-14}$  			&$1.2\pm0.6$	&-\\
\hline
\hline
\end{tabular}
\end{center}
\end{table*}

Experimentally, $D^{+(0)}$ decays have been studied intensively in the past two decades~\cite{PDGweb}.
However, existing measurements of charmed-strange meson decays suffer from poor precision due to the limited size of available data samples and a relatively small production cross section in $\ee$ annihilation~\cite{cleoxsec}.
The most recent measurement of $\br{}(\Dsp\ra \ksz \kp)=(1.52\pm0.05_{\rm stat.}\pm0.03_{\rm syst.})\%$ was reported by the CLEO Collaboration~\cite{cleo-c:res}; the result was obtained using a global fit to  multiple decay modes reconstructed in an $\ee$ annihilation sample corresponding to an integrated luminosity of  586\,$\ipb$ at a center-of-mass energy $\sqrt{s} = 4.17\,\gev$.
The Belle Collaboration reported a measurement of the branching fraction  $\br{}(\Dsp\ra  \kzbar \kp)$ (ignoring the contribution from $\kz K$)~\cite{belle:res} using a data sample corresponding to an integrated luminosity of 913~$\ifb$ collected at $\sqrt{s}$  around the $\Upsilon(4S)$ and $\Upsilon(5S)$ resonances. Neither $\br{}(\Dsp\ra \klz\kp)$ nor the corresponding $\ksz$-$\klz$ asymmetry have been measured yet.

In this paper, measurements of the absolute branching fractions for the decays $\Dsp\ra\pksk$ and $\Dsp\ra\klz\kp$, the $\ksz$-$\klz$ asymmetry, and the corresponding $CP$ asymmetries are performed using a sample of  $\ee$ annihilation data collected at $\sqrt{s} = 4.178~\gev$ with the BESIII detector at the BEPCII.  The data sample corresponds to an integrated luminosity of  3.19~$\ifb$. Throughout the paper, charge conjugation modes are implicitly implied, unless otherwise noted.

\section{BESIII DETECTOR AND MONTE CARLO SIMULATION}
\label{detmc}
The BESIII detector is a magnetic spectrometer that operates at the BEPCII $\ee$ collider~\cite{bepcii}. The detector has a cylindrical geometry that covers 93\% of the 4$\pi$ solid angle and consists of several subdetectors.  A main drift chamber (MDC) with 43 layers surrounding the beam pipe measures momenta and specific ionization of charged particles.  Plastic scintillator time of flight counters (TOF), located outside of the MDC, provide charged-particle identification information, and an electromagnetic calorimeter (EMC), consisting of 6240 CsI(Tl) crystals, detects electromagnetic showers. These subdetectors are immersed in a magnetic field of 1~T, produced by a superconducting solenoid, and are surrounded by a multi-layered resistive-plate chamber (RPC) system (MUC) interleaved in the steel flux return of the solenoid, providing muon identification. In 2015, BESIII was upgraded by replacing the two endcap TOF systems with multi-gap RPCs, which achieve a time resolution of 60~ps ~\cite{ETOF}. A detailed description of the  BESIII detector is presented in Ref.~\cite{Ablikim:2009aa}.

The performance of the BESIII detector is evaluated using a {\sc geant4}-based~\cite{geant4}  Monte Carlo (MC) program that includes a description of the detector geometry as well as simulating its response. 
In the MC simulation, the production of open charm processes directly produced via $e^+e^-$  annihilation are modeled with the generator {\sc conexc}~\cite{conexc}, which includes the effects of the beam energy spread and initial-state radiation (ISR). The ISR production of vector charmonium states ($\psi(3770)$, $\psi(3686)$ and $J/\psi$) and the continuum processes ($q\bar{q}$, $q=u,d,s$) are incorporated in {\sc kkmc}~\cite{kkmc}. The known decay modes are generated using {\sc evtgen}~\cite{evtgen}, which assumes the branching fractions reported in Ref.~\cite{PDGweb}; the fraction of unmeasured decays of  charmonium states is generated with {\sc lundcharm}~\cite{lundcharm}. The final-state radiation (FSR) from charged tracks is simulated by the {\sc photos} package~\cite{photos}. A generic MC sample with equivalent luminosity 35 times that of data is generated to study the background. It contains open charm processes, the ISR return to charmonium states at lower mass, and continuum processes (quantum electrodynamics and $q\bar{q}$).  The signal MC samples of 5.2 million $\ee\ra D_{s}^{*\pm} D_{s}^{\mp} $ events are produced; in these samples the $D_{s}^{*\pm}$ decays into $\gam/\piz/\ee D_{s}^{\pm}$, while one $\Ds$ decays into a specific mode in Table~\ref{tab:NST_effSTDT} and the other into the final states of interest $\ksz\kpm$ or $\klz\kpm$. The signal MC samples are  used to determine the distributions of kinematic variables and estimate the detection efficiencies.

\section{DATA ANALYSIS}
The cross section to produce $\ee\ra D_{s}^{*\pm} D_{s}^{\mp}$ events at $\sqrt{s}=4.178~\gev$ is ($889\pm59_{\rm stat.}\pm47_{\rm syst.} )$ pb, which is one order of magnitude larger than that to produce $\ee\ra\DspDsm$ events~\cite{cleoxsec}. Furthermore, the decay branching fraction $\br{}(\Dsstp\ra\gamma\Dsp)$ is $(93.5\pm0.7)\%$ ~\cite{PDGweb}. Therefore, in the data sample used, $\Dsp$ candidates arise mainly from the process $\ee\ra D_{s}^{*\pm} D_{s}^{\mp}\ra\gamma\DspDsm$, along with small fractions from the processes $\ee\ra D_{s}^{*\pm} D_{s}^{\mp}\ra\piz\DspDsm$ and $\ee\ra\DspDsm$.  The outline of the reconstruction is described first, with all details given later in this section.

In this analysis, a sample of $\Dsm$ mesons is reconstructed first, which are referred to as ``single tag (ST)'' candidates.  The ST candidates are reconstructed in 13 hadronic decay modes that are listed in Table~\ref{tab:NST_effSTDT}. The $\Dsm\ra\mksk$ tag mode is not included to avoid double counting in $\Dsp\ra\pksk$ measurement. Here, $\piz$ and $\eta$ candidates are reconstructed from a pair of photon candidates, $\ksz$ candidates are formed from $\pip\pim$ pairs, and $\rho^{\pm(0)}$ candidates are reconstructed from $\pipm\pi^{0(\mp)}$ pairs, unless otherwise indicated by a subscript. 
\begin{table}[htbp]
\caption{Summary of the $\Dsm$ ST yields, along with the ST and DT detection efficiencies for that decay mode. The uncertainty is statistical only. The decay branching fractions of subsequent decays in ST side are not included in the efficiencies. The decay branching fraction of $\ksz\ra\pip\pim$ in signal side is included in $\epsilon_{\rm DT}^{\ksz}$.
}
\label{tab:NST_effSTDT}
\footnotesize
\begin{center}
\begin{tabular}{l|cccc }
\hline
\hline
Tag mode        		&   $N_{\rm ST}$ 		& $\epsilon_{\rm ST}$($\%$)	& $\epsilon_{\rm DT}^{\ksz}$($\%$)		& $\epsilon_{MM^{2}}^{\klz}$($\%$)\\
\hline
$\mkkpi$        		&141285$\pm$631      	&42.15$\pm$0.03       	&13.58$\pm$0.07				&16.33$\pm$0.10\\
$\mkpipi$       		&\phantom{0}18051$\pm$575     	&48.84$\pm$0.26     	&16.35$\pm$0.08				&19.73$\pm$0.12\\
$\mpipipi$      		&\phantom{0}40573$\pm$964       	&56.05$\pm$0.18      	&18.47$\pm$0.08				&22.55$\pm$0.12\\
$\mkkpipiz$     		&\phantom{0}41001$\pm$840       	&10.61$\pm$0.03      	&\phantom{0}3.86$\pm$0.04		&\phantom{0}5.02$\pm$0.06\\
$\mpietaprhog$  		&\phantom{0}26360$\pm$833       	&35.33$\pm$0.16    	&12.41$\pm$0.07				&15.59$\pm$0.10\\
$\rho^{-}\eta$    		&\phantom{0}32922$\pm$878      	&16.65$\pm$0.06     	&\phantom{0}5.99$\pm$0.06		&\phantom{0}8.84$\pm$0.09\\
$\mkskpipi$     		&\phantom{00}8081$\pm$283        	&18.47$\pm$0.11      	&\phantom{0}6.16$\pm$0.05		&\phantom{0}7.72$\pm$0.07\\
$\mksktpi$      		&\phantom{0}15331$\pm$249       	&21.44$\pm$0.06    	&\phantom{0}6.82$\pm$0.05		&\phantom{0}8.21$\pm$0.07\\
$\mkskpiz$      		&\phantom{0}11380$\pm$385       	&16.97$\pm$0.12    	&\phantom{0}5.94$\pm$0.05		&\phantom{0}7.82$\pm$0.07\\
$\mkskspi$      		&\phantom{00}5015$\pm$164        	&22.86$\pm$0.11    	&\phantom{0}6.95$\pm$0.05		&\phantom{0}8.98$\pm$0.07\\
$\pim\eta$     		&\phantom{0}19050$\pm$512       	&46.60$\pm$0.19      	&16.06$\pm$0.07				&21.99$\pm$0.13\\
$\pim\etap_{\pip\pim\eta}$&\phantom{00}7694$\pm$137       	&18.80$\pm$0.05  		&\phantom{0}6.16$\pm$0.05		&\phantom{0}8.45$\pm$0.08\\
$\mpietathpi$   		&\phantom{00}5448$\pm$169        	&22.30$\pm$0.11    	&\phantom{0}7.47$\pm$0.06		&\phantom{0}9.70$\pm$0.08\\
\hline
\hline
\end{tabular}
\end{center}
\end{table}

In the sample of events with an ST candidates, the process $\Dsp\ra\ksz\kp$ is reconstructed by selecting a charged kaon and a $\ksz$ candidates from those not used to reconstruct the ST candidates, which is referred as ``double tag (DT)''. To reconstruct the $\Dsp\ra\klz\kp$ decay, the photon from the decay $\Ds^{*\pm}\ra\gamma\Ds^{\pm}$ and the charged kaon from $\Dsp$ decay  are selected to determine the missing-mass-squared
\begin{eqnarray}
  MM^{2} =  (P_{\ee}-P_{\Dsm}-P_{\gam}-P_{\kp})^{2}, \label{eq:MM2}
\end{eqnarray}
where $P_{\ee}$ is the four-momentum of the $\ee$ initial state and $P_{i}~(i=\Dsm,\gam,\kp)$ is the four-momentum of the corresponding particle.

Ignoring the small contribution from the process $\ee\ra\DspDsm$, the numbers of ST ($N_{\rm ST}^i$) and DT ($N_{\rm DT}^i$) events, for a specific tag mode $i$, are
\begin{eqnarray}
 &&  N_{\rm ST}^i = 2\times N_{D_{s}^{*\pm}D_{s}^{\mp}}\times\br{\rm tag}^i\times\epsilon_{\rm ST}^{i}, \label{eq:NST}\\ 
 &&  N_{\rm DT}^i = 2\times N_{D_{s}^{*\pm} D_{s}^{\mp}}\times\br{\rm tag}^i\times\br{\rm sig}\times\epsilon_{\rm DT}^i,\label{eq:NDT}
\end{eqnarray}
\noindent
respectively. Here, $N_{D_{s}^{*\pm} D_{s}^{\mp}}$ is the total number of $\ee\ra D_{s}^{*\pm} D_{s}^{\mp}$ events in the data sample,
$\br{\rm tag}^i$ is the branching fraction for the $i^{\rm th}$ ST decay mode,
$\br{\rm sig}$ is the branching fraction of the signal decay;
$\epsilon_{\rm ST}^i$ and $\epsilon_{\rm DT}^i$ are the ST and DT detection efficiencies, respectively, which are evaluated from the signal MC samples corresponding to the $i^{\rm th}$ tag mode. The value of $\epsilon_{\rm DT}^i$ includes the branching fraction $\br{}(\ksz\ra\pip\pim)$ of signal side in the analysis of $\Dsp\ra\ksz\kp$.
The factors of two in Eqs.~(\ref{eq:NST}) and  (\ref{eq:NDT}) are the result of including charge-conjugated modes in the analysis.
We combine  Eqs.~(\ref{eq:NST}) and (\ref{eq:NDT}) for each of the 13 tag modes to obtain  
\begin{eqnarray}
\br{\rm sig} = \frac{N_{\rm DT}^{\rm tot}}{\sum\limits_{i}N_{\rm ST}^{i}\times\epsilon_{\rm DT}^{i}/\epsilon_{\rm ST}^{i}},
\label{eq:BF}
\end{eqnarray}
\noindent
where $N_{\rm DT}^{\rm tot}=\sum\limits_{i}N_{\rm DT}^{i}$ is the total number of DT events.

\subsection{Selection of ST events}\label{STEVTSEL}
Good charged tracks, except for the daughter tracks of $\ksz$ candidates, are selected by requiring  the track trajectory approaches the interaction point (IP) within $\pm 10$~cm along the beam direction and within 1~cm in the plane perpendicular to the beam direction. In addition, the polar angle $\theta$ between the direction of the charged track and the beam direction must be within the detector acceptance by requiring $|\cos\theta |<0.93$. Charged particle identification (PID) is performed by combining the ionization-energy loss $(\mathrm{d}E/\mathrm{d}x)$ measured by the MDC and the time-of-flight measured by the TOF system. Each charged track is characterized by the PID likelihood for the pion and kaon hypotheses, which are ${\cal{L}}(\pi)$ and ${\cal{L}}(K)$, respectively. A pion [kaon] candidate is identified if it satisfies the condition ${\cal{L}}(\pi) > {\cal{L}}(K)$  $[{\cal{L}}(K) > {\cal{L}}(\pi)]$.

Good photon candidates are selected from isolated electromagnetic showers which have a minimum energy of 25~MeV in the EMC barrel region ($|\cos\theta | < 0.8$) or 50~MeV in the EMC endcap region ($0.86 < | \cos\theta | < 0.92$). To reduce the number of photon candidates that result from noise and beam backgrounds, the time of the shower measured by the EMC is required to be less than 700~ns after the beam collision. The opening angle between a photon and the closest charged track is required to be greater than 10\degree, which is used to remove electrons, hadronic showers and photons from FSR. $\piz$ and $\eta\ra\gamma\gamma$ candidates are reconstructed from pairs of photon candidates that have an invariant mass within the intervals (0.115, 0.150) and (0.50, 0.57)~$\gevcc$, respectively. To improve the momentum resolution, a kinematic fit is performed, constraining the $\gam\gam$ invariant mass to its nominal value~\cite{PDGweb}; the $\chi^2$ of the fit is required to be less than 20 to reject combinatorial background. $\eta\ra\pip\pim\piz$ candidates are selected by requiring the corresponding invariant mass to be within the interval (0.534, 0.560)~$\gevcc$.

In order to improve the efficiency of the $\ksz$ selection, $\ksz$ candidates are reconstructed from tracks assumed to be pions without PID, and the daughter tracks are required to have a trajectory that approaches the IP to within $\pm 20$~cm along the beam direction and $|\cos\theta | < 0.93$. The $\ksz$ candidates are formed by performing a vertex-constrained fit to all oppositely-charged track pairs. To suppress combinatorial background, the $\chi^2$ of the vertex fit is required to be less than 200 and a secondary vertex fit is performed to ensure that the $\ksz$ candidate originates from the IP.  The flight length $L$, defined as the distance between the common vertex of the $\pip\pim$ pair and the IP in the plane perpendicular to beam direction,  is obtained in the secondary vertex fit, and is required to satisfy $L > 2\sigma_{L}$, where $\sigma_{L}$ is the estimated uncertainty on $L$; this criterion removes combinatorial background formed from tracks originating from the IP.  The four-momenta after the secondary vertex fit are used in the subsequent analysis. The $\ksz$ candidate is required to have a mass within the interval (0.487, 0.511)~$\gevcc$.

$\etap$ candidates are reconstructed via the decay modes $\gam\rho^0$ and $\pip\pim\eta$ by requiring the corresponding invariant masses to be within the intervals (0.936, 0.976) and (0.944, 0.971)~$\gevcc$, respectively. The $\rho^0$ candidates are reconstructed from $\pip\pim$ pairs that have a mass greater than 0.52~$\gevcc$. The $\rho^{\pm}$ candidates are reconstructed from $\pipm\piz$ combinations that have an invariant mass within the interval (0.62, 0.92)~$\gevcc$.

 To suppress the background with $\Dst$ decay $\Dst\ra\pi D$, the momentum of charged and neutral pions is required to be greater than 100~$\mevc$. For $\mkpipi$ ST candidates, the invariant mass of the $\pip\pim$ pair is required to be outside the interval (0.478, 0.518) $\gevcc$ to remove  $\Dsm\ra\ksz\km$ decays. The ST $\Dsm$ candidates are reconstructed via all the possible selected particles combinations.  
  \begin{figure*}[htbp]
    \begin{center}
    \includegraphics[width=1.0\textwidth,height=0.7\textwidth]{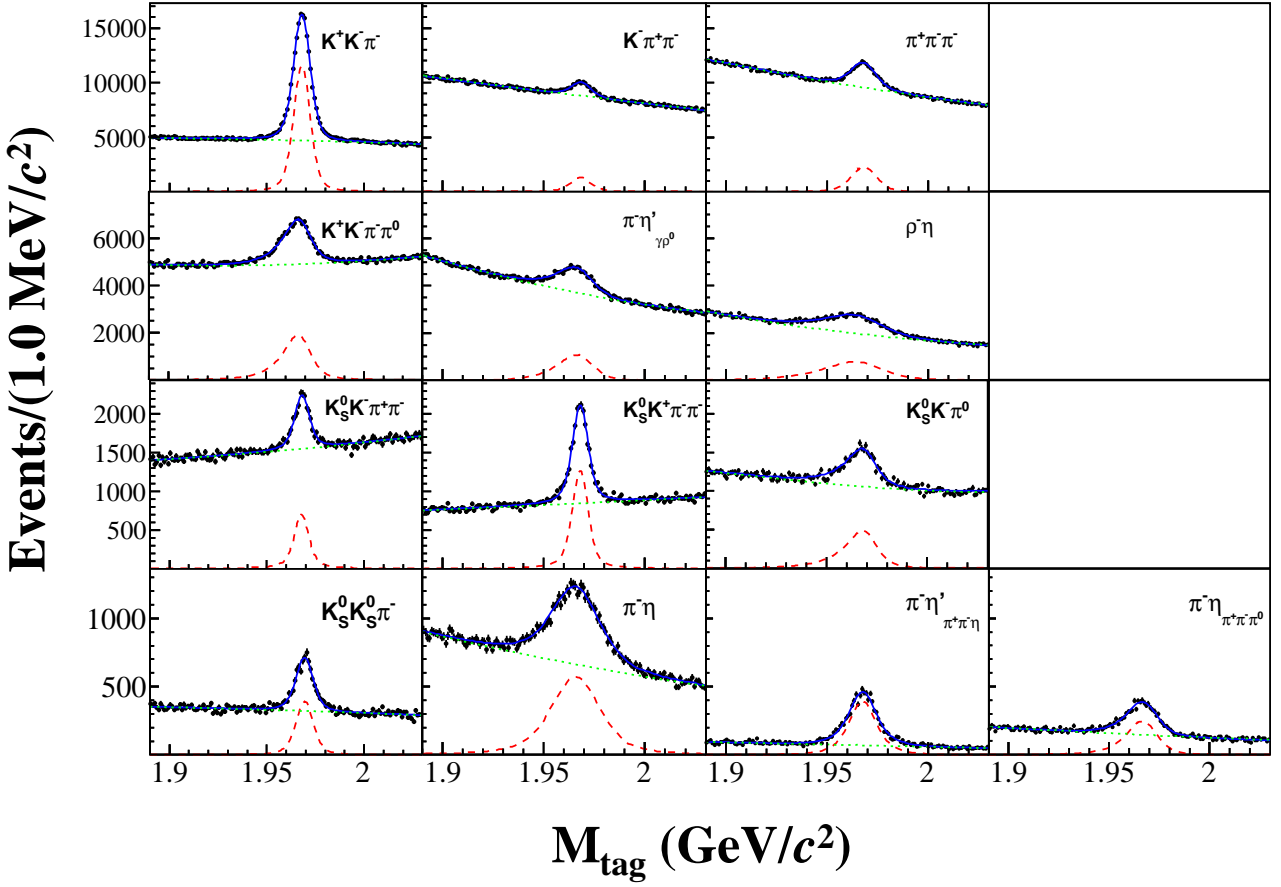}
    \caption{Fits to $M_{\rm tag}$  distributions for each ST mode. The dots with error bars are data, the blue solid curves are  the overall fit results, the red dashed curves are the signal, and the green dotted curves are the background.}
    \label{fig:STfit}
    \end{center}
    \end{figure*}
    
The invariant mass of the system recoiling against the selected $\Dsm$ is defined as 
\begin{eqnarray}
M_{\rm rec}=\sqrt{(\sqrt{s}-\sqrt{p^2+M_{\Ds}^2})^2-p^2},
\label{eq:Mrec}
\end{eqnarray}
\noindent
where $p$ is the momentum of the ST $\Dsm$ candidate in $\ee$ CM frame, and $M_{\Ds}$ is the nominal mass of the $\Ds$  meson~\cite{PDGweb}. $M_{\rm rec}$  is required to be within the interval (2.05, 2.18) $\gevcc$. For a specific ST mode, if there are multiple combinations satisfying the selection criteria, only the candidate with the minimum value of $|M_{\rm rec}-M_{\Dsst}|$ is retained for further analysis.  These requirements also accept the events in which the ST $\Ds$ comes from the decay of the primary $\Dsst$.

To determine the ST yield, a binned maximum likelihood fit to the distribution of the $\Dsm$ invariant mass $M_{\rm tag}$ is performed for each tag mode; the distributions and fit results are shown in Fig.~\ref{fig:STfit}. In the fit, the probability density function (PDF) that describes the signal is the shape of the signal MC distribution, taken as a smoothed histogram and convolved with a Gaussian function to account for any resolution difference between data and MC simulation. The background is described by a second or third-order Chebychev polynomial function. The ST yields determined by the fits, along with the corresponding $\epsilon_{\rm ST}^i$ estimated from the generic MC sample, are summarized in Table~\ref{tab:NST_effSTDT}.

\subsection{\boldmath Branching fraction measurement of $\Dsp\ra\pksk$} \label{KSKMEAS}
 The signal decay $\Dsp\ra \pksk$  is reconstructed recoiling against the selected ST $\Dsm$ candidate.
 We select a  $\Dsp\ra \pksk$ candidate if there is only one $\ksz$ candidate and one good track, which is identified as a kaon and has  positive charge, recoiling against the ST $\Dsm$ candidate;  $K^{+}$ and $\ksz$ candidates are selected by applying the selection criteria described in Sec.~\ref{STEVTSEL}. In addition, to suppress combinatorial backgrounds, we reject events in which there are additional charged tracks that satisfy $|\cos\theta|<0.93$ and approach the IP along the beam direction within $\pm20$~cm.

To determine the DT signal yield, a two-dimensional (2D) unbinned maximum likelihood fit is performed on the invariant mass of the $\ksz$ and $K^{+}$ $(M_{\pksk})$ versus $M_{\rm tag}$ distribution of selected events, which is summed over the 13 ST modes, as shown in Fig.~\ref{fig:Ks2D}. In the fit, the total PDF is described by summing over the individual PDFs for the following signal and background components, where $x$ represents $M_{\pksk}$, and $y$ stands for $M_{\rm tag}$.
\begin{figure}[htbp]
\begin{center}
\begin{overpic}[width=0.46\textwidth,height=0.32\textwidth]{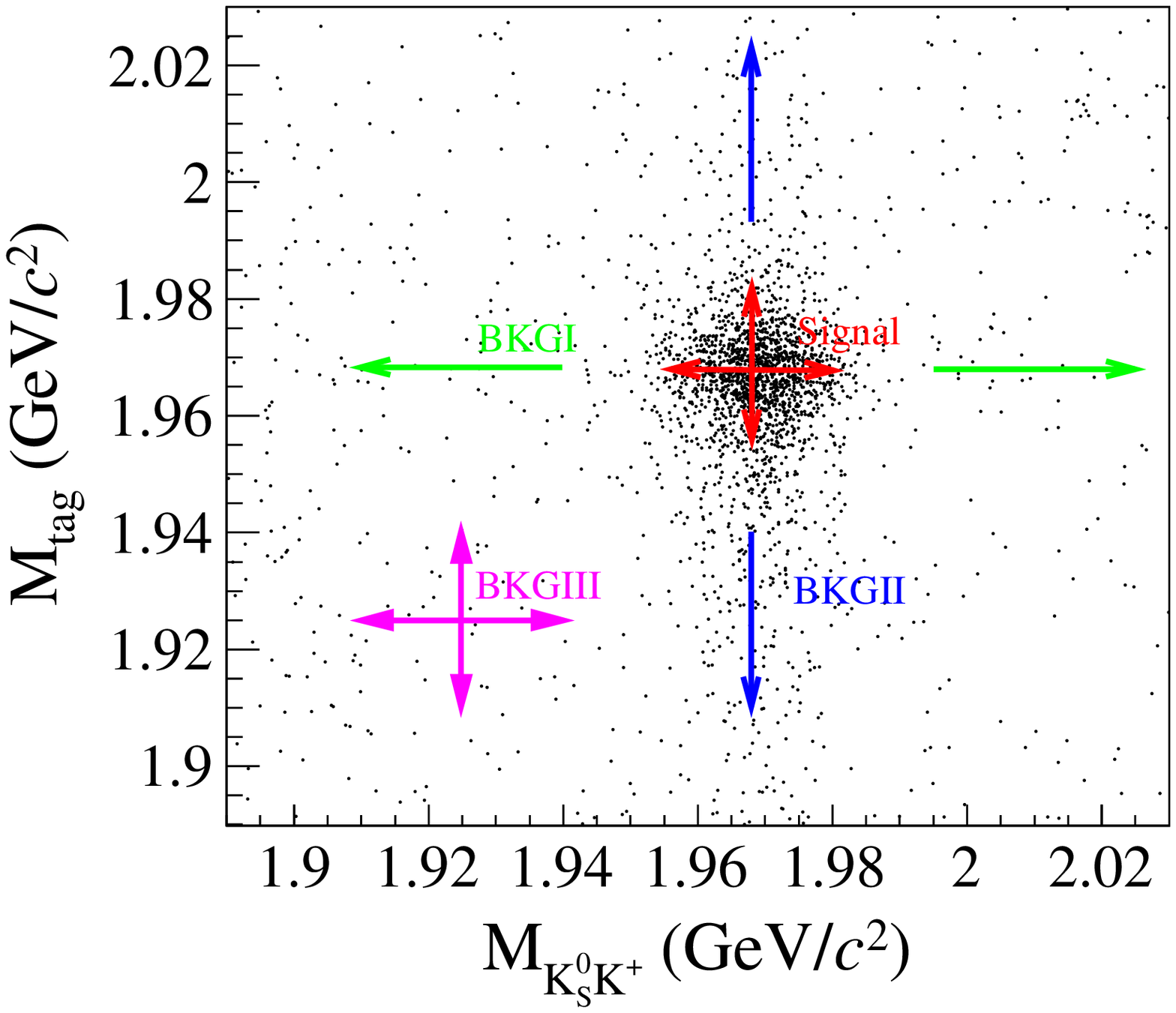}
\end{overpic}
\caption{Distribution of  $M_{\rm tag}$ vs. $M_{\pksk}$  for $\Dsp\ra\ksz\kp$ candidates in data, summed over the 13 tag modes.}
\label{fig:Ks2D}
\end{center}
\end{figure}

\begin{itemize}
\item Signal: ~$F_{\rm sig}(x,y)$ $\otimes$ $G(x;\mu_{x},\sigma_{x})$ $\otimes$ $G(y;\mu_{y},\sigma_{y})$

	  $F_{\rm sig}(x, y)$ is a 2D function derived from the signal MC distribution by using a smoothed 2D histogram; $G(x;\mu_{x},\sigma_{x})$ and $G(y;\mu_{y},\sigma_{y})$ are Gaussian functions that compensate for any resolution difference between data and MC simulation for the variables $M_{\pksk}$ and $M_{\rm tag}$, respectively. In the 2D fit, the parameters of $G(x;\mu_{x},\sigma_{x})$ and $G(y;\mu_{y},\sigma_{y})$ are fixed to the values determined by fitting the corresponding one-dimensional (1D) distributions.

\item BKGI: ~$F_{\rm BKGI}(x,y)$ $\otimes$ $G(y;\mu_{y},\sigma_{y})$

  This PDF describes the background composed of a correctly reconstructed ST $\Dsm$ recoiling against combinatorial background, which are distributed in the horizontal band in Fig.~\ref{fig:Ks2D}. $F_{\rm BKGI}(x,y)$ is derived from the distribution of this type of background in the generic MC sample by using a kernel density estimation method (KEYS)~\cite{Cranmer:2000du}. The resolution function $G(y;\mu_{y},\sigma_{y})$ is the same as that in the Signal PDF.

\item BKGII: ~$F_{\rm BKGII}(x,y)$ $\otimes$ $G(x;\mu_{x},\sigma_{x})$

 This PDF describes the background composed of an incorrectly reconstructed ST $\Dsm$ recoiling against a correctly reconstructed signal candidate, which are distributed in the vertical band in  Fig.~\ref{fig:Ks2D}.
 $F_{\rm BKGII}(x,y)$ is derived from the distribution of this type of background in the generic MC sample by using KEYS. The resolution function $G(x;\mu_{x},\sigma_{x})$ is the same as that in the Signal PDF.

\item BKGIII: ~$P_{\rm BKGIII}(x)$ $\times$ $P_{\rm BKGIII}(y)$

This PDF describes the combinatorial background composed of events in which neither the ST $\Dsm$ nor signal $\Dsp$ candidate is correctly reconstructed. These background events do not have any peaking components in either variable. Therefore, BKGIII events are described by two independent  second-order polynomials, $P_{\rm BKGIII}(x)$ and $P_{\rm BKGIII}(y)$, with their parameters determined by the fit to data.
\end{itemize}

The 2D fit gives a signal yield of $1782~\pm~47$, where the uncertainty is statistical. The $M_{\pksk}$ and $M_{\rm tag}$ distributions for the data, with the projections of the fit results superimposed, are shown in Fig.~\ref{fit:Ks2Dfit}.
The corresponding DT detection efficiencies for the individual ST mode, obtained with the signal MC samples, are summarized in  Table~\ref{tab:NST_effSTDT}. Using Eq.~(\ref{eq:BF}), the branching fraction is determined to be $\br{}(\Dsp\ra\pksk)=(1.425~\pm~0.038_{\rm stat.})\%$.

\begin{figure}[htbp]
\begin{center}
\begin{overpic}[width=0.5\textwidth,height=0.36\textwidth]{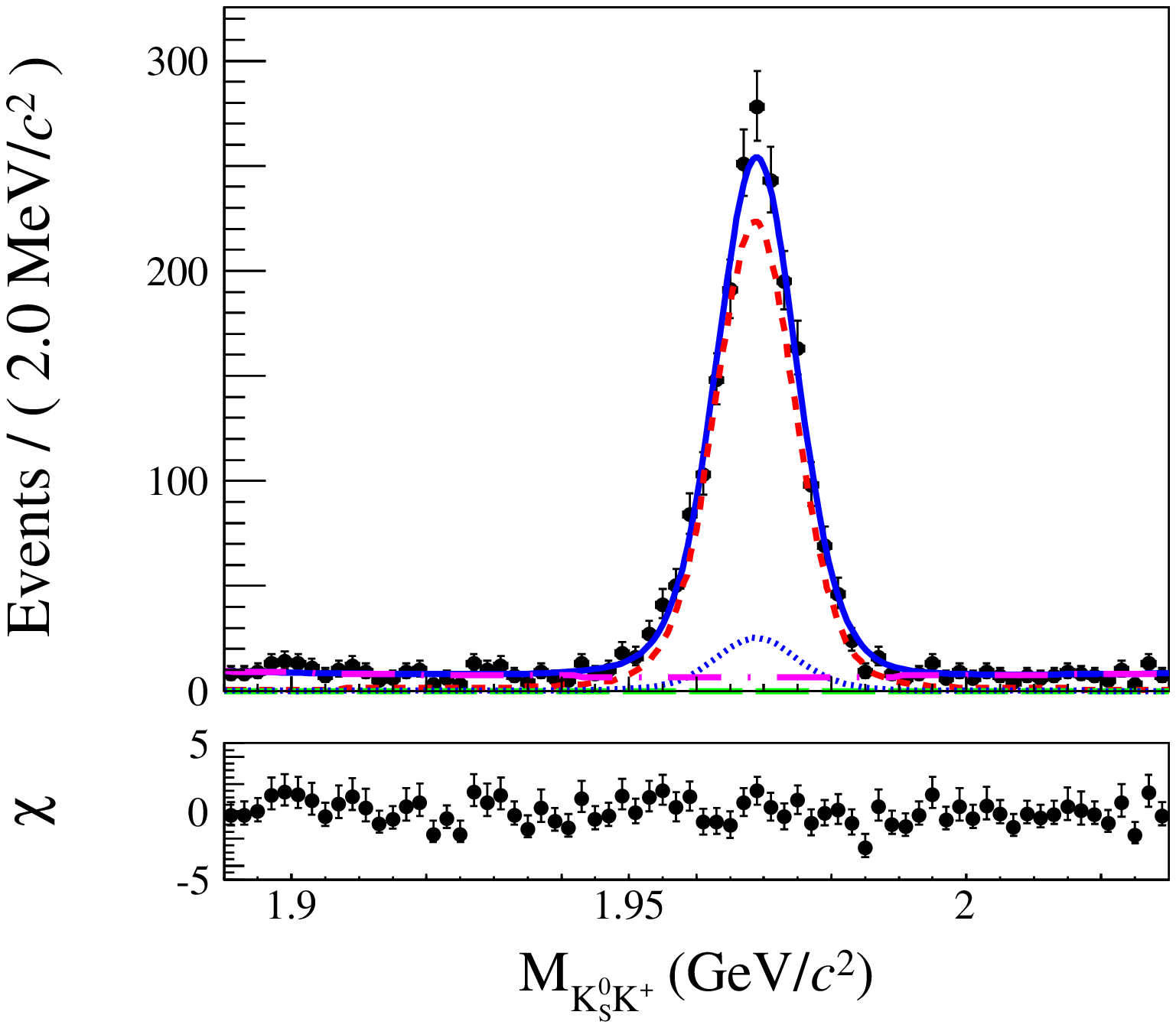}
  \put(70,62){\boldmath{(a)}}
\end{overpic}
\begin{overpic}[width=0.5\textwidth,height=0.36\textwidth]{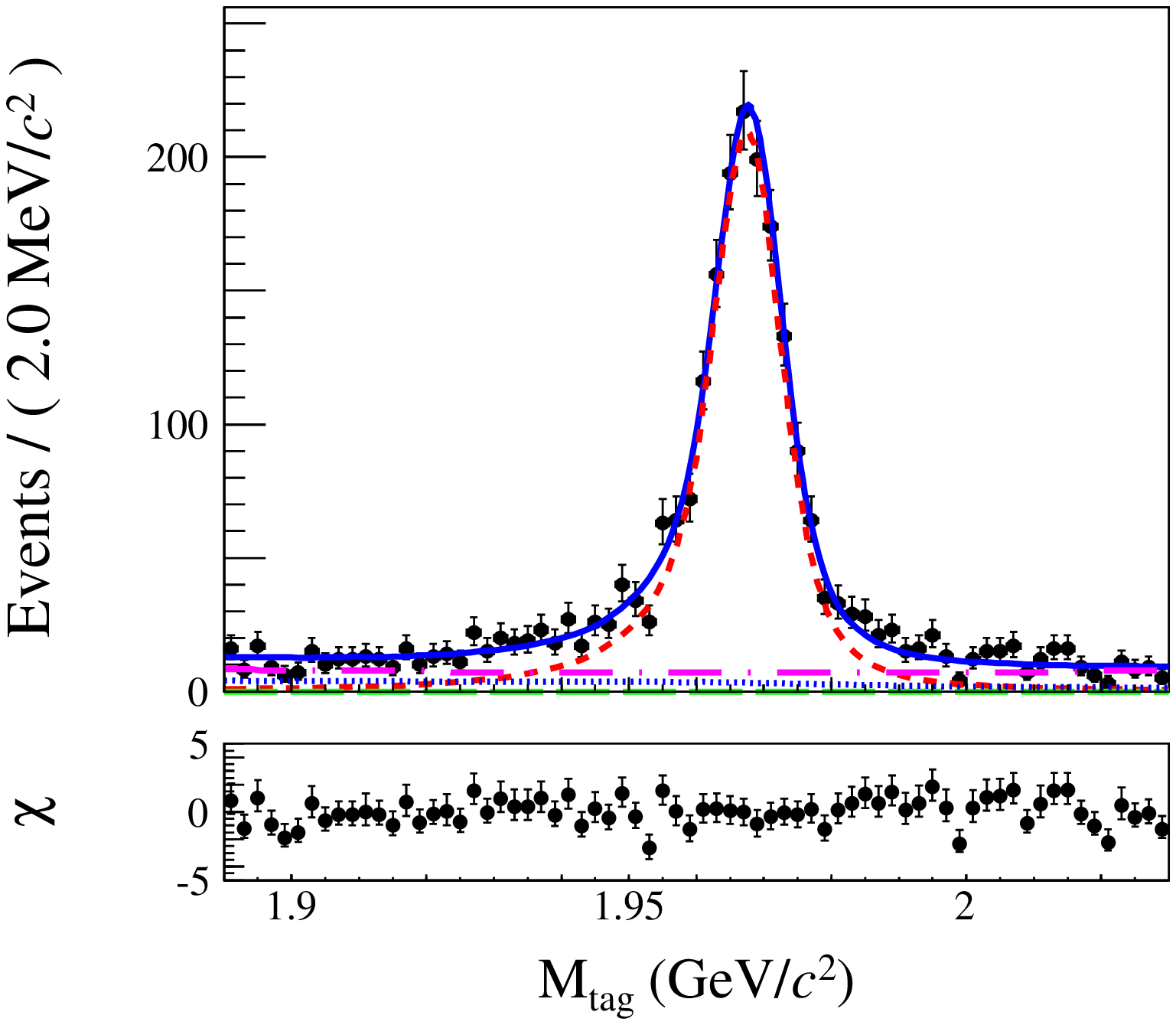}
  \put(70,62){\boldmath{(b)}}
\end{overpic}
\caption{(a) Distributions of $M_{\pksk}$ and (b) $M_{\rm tag}$, summed over the 13 tag modes, with the projection of the fit result superimposed. The data is shown as the black dots with error bars, the blue solid line is the total fit projection, the red short-dashed line is the projection of the signal component, the green long-dashed line is the projection of the BKGI component, the blue dotted line is the projection of the  BKGII component, and the magenta dot-dashed line is the projection of the BKGIII component. The residual $\chi$ between the data and the total fit result, normalised by the uncertainty, is shown beneath the figures.}
\label{fit:Ks2Dfit}
\end{center}
\end{figure}

\subsection{\boldmath Branching fraction measurement of $\Dsp\ra\klz\kp$}\label{KLKMEAS}
The $\Dsp\ra \pklk$ candidates are reconstructed by requiring  the event has only one good track recoiling against the  ST $\Dsm$ candidate; the charged track is required to be identified as a kaon and have opposite charge compared with ST $\Dsm$. The $K^{+}$ is selected with the criteria described in Sec.~\ref{STEVTSEL}. We further suppress combinatorial backgrounds by requiring no additional charged tracks that statisfy the requirements described in Sec.~\ref{KSKMEAS}.

   In this analysis, the ST and signal candidates are assumed to originate from the decay chain $\ee\ra D_{s}^{*\pm} D_{s}^{\mp}\ra\gamma\Dsp\Dsm$, with one $\Dsm$ decaying into any of ST modes, and the other decaying into $\klz\kp$. We reconstruct the  $\klz$ candidate using a kinematic fit that applies constraints arising from the masses of the ST $\Dsm$ candidate, the signal $\Dsp$ candidate, the intermediate state $D_{s}^{*\pm}$, and the initial four-momenta of the event.  In the kinematic fit, the $\klz$ signal candidate is treated as a missing particle whose four-momentum is determined by the fit. The fit is performed to select $\gamma$ candidate from the decay $D_{s}^{*\pm}\ra\gamma\Dspm$  under two different hypotheses that constrain either the invariant mass of the selected $\gamma$ and signal $\Dsp$ or the selected $\gamma$ and the ST $\Dsm$ to  the nominal mass of the $\Dsstm$ meson; the hypothesis that results in the minimum value of $\chi^2$ is assumed to be the correct topology. If there are multiple photon candidates,  which are not used to reconstruct the ST candidate, the fit is repeated for each candidate and the photon that results in the minimum value of the $\chi^2$ is retained for further analysis. For each event, the four-momentum of the missing particle assumed in the kinematic fit is used to determine the $MM^{2}$ of the $\klz$ candidate. In order to reduce combinatorial background, $\chi^{2} < 40$ is required.  To further suppress background with multiple photons, we reject those events with additional photons which have an energy larger than 250~$\mev$ and an opening angle with respect to the direction of missing particle greater than $15^{\circ}$. 

\begin{figure}[htbp]
\begin{center}
\begin{overpic}[width=0.5\textwidth,height=0.36\textwidth]{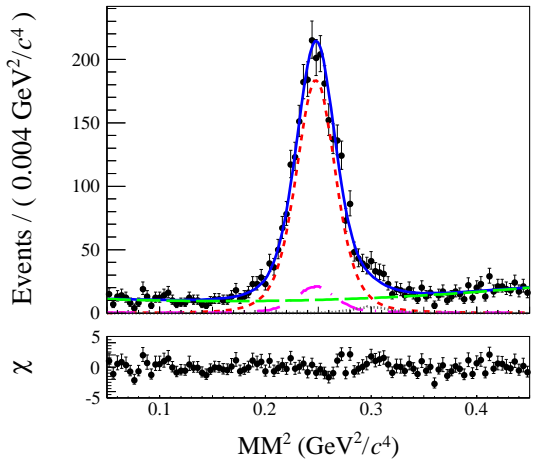}
\end{overpic}
\caption{Distribution of $MM^{2}$ summed over 13 tag modes with the fit result superimposed. The data is shown as the dots with error bars, the blue solid line is the total fit result, the red short-dashed line is the signal component of the fit, the magenta dot-dashed line is the component of the peaking background from $\Dsp\ra\pksk$ decays and the grey dotted line is the component of the peaking background from $\Dsp\ra\eta\kp$ decays, the green long-dashed line is the non-peaking background component. The residual $\chi$ between the data and the total fit result, normalised by the uncertainty, is shown beneath the figures.}
\label{fig:KLfit}
\end{center}
\end{figure}

  To determine the signal yield, an unbinned maximum likelihood fit is performed on the $MM^{2}$ distribution of selected events from all 13 ST modes combined, as shown in Fig.~\ref{fig:KLfit}. In the fit, three components are included: signal, peaking, and non-peaking backgrounds. The PDFs of these components are described below, where $x$ represents $MM^{2}$.
\begin{itemize}
\item Signal: $F_{\rm sig}(x)$ $\otimes$ $G(x;\mu'_{x},\sigma'_{x})$

  $F_{\rm sig}(x)$ is derived from the signal MC distribution as a smoothed histogram, and $G(x;\mu'_{x},\sigma'_{x})$ is a Gaussian function that accounts for any resolution difference between data and MC simulation. The value of $\sigma'_x$ is fixed in the data fit to the value obtained from a fit to the $MM^{2}$ distribution obtained from a $\Dsp\ra\pksk$ control sample where the $\ksz$ is ignored in the reconstruction.

\item Peaking background:  $F_{\rm bkg}^{\ksz(\eta)}(x)$ $\otimes$ $G(x;\mu'_{x},\sigma'_{x})$

    $F_{\rm bkg}^{\ksz(\eta)}(x)$ is derived from the distribution of $\Dsp\ra\pksk$ $(\Dsp\ra\eta\kp)$ MC simulated events by using a smoothed histogram.  These events form a peaking background if the $\ksz$ or $\eta$ is not reconstructed.  Here, $G(x;\mu'_{x},\sigma'_{x})$ is the Gaussian resolution function, whose parameters are the same as those used in the signal PDF. The expected yields of $\Dsp\ra\pksk$ and $\Dsp\ra\eta\kp$ are fixed to 263 and 57, respectively. The expected peaking background yields are estimated by using the equation $N_{MM^{2}}^{\rm data}=N_{\rm DT}^{\rm data}\times\epsilon_{MM^{2}}^{\rm MC}/\epsilon_{\rm DT}^{\rm MC}$, where $N_{MM^{2}}^{\rm data}$ is the number of expected peaking background events and $N_{\rm DT}^{\rm data}$ is the yield of $\Dsp\ra\pksk$ or $\Dsp\ra\kp\eta$ selected by using the DT method. Here, $\epsilon_{MM^{2}}^{\rm MC}$ and $\epsilon_{\rm DT}^{\rm MC}$ are the detection efficiencies of the nominal analysis and the DT method for each mode, respectively; these are estimated from MC simulation samples. The uncertainties of estimated event numbers  for $\Dsp\ra\pksk$ and $\Dsp\ra\eta\kp$ are 19 and 12, which will be used in systematic uncertainty study.
\item Non-peaking background:  $P(x)$

$P(x)$ is a function to describe the combinatorial background, which is not expected to peak in the $MM^{2}$ distribution. $P(x)$ is a second-order polynomial function whose parameters are determined from the fit to data.
\end{itemize}

The fit to the $MM^{2}$ distribution is shown in Fig.~\ref{fig:KLfit}. The signal yield determined by the fit is $2349~\pm~61$ events, where the uncertainty is statistical. Using Eq.~(\ref{eq:BF}), the branching fraction is calculated to be $\br{}(\Dsp\ra\klz\kp)=(1.485~\pm~0.039_{\rm stat.})\%$, where the DT detection efficiencies $\epsilon_{MM^{2}}^{\klz}$ used are summarized in Table~\ref{tab:NST_effSTDT}; the values of $\epsilon_{MM^{2}}^{\klz}$ are estimated from signal MC samples.

\begin{table*}[ht]
\caption{Summary of relative systematic uncertainties (\%) of the branching fraction measurements and the absolute systematic uncertainties (\%) of the $A_{\rm CP}$ and $R(\Dsp)$ measurements.}
\label{tab:sys}
\footnotesize
\begin{center}
\begin{tabular}{lcc|cccc}
\hline
\hline
Source   &$\br{}(\Dsp\ra\pksk)$  &$\br{}$($\Dsp\ra\klz \kp)$ &$R(\Dsp\ra \kz_{S,L} \kp)$ & $A_{\rm CP}(\Dspm\ra\ksz\kpm)$ & $A_{\rm CP}(\Dspm\ra\klz\kpm)$\\
\hline
$\kp/\km$ tracking    				&     0.5 	&    0.5    	&    -    	&	 0.4   	&	0.4 \\
$\kp/\km$ PID         				&     0.5     	&    0.5    	&    -    	&	 0.4   	&   	0.4\\
$\ksz$ reconstruction 				&     1.5    	&     -     	&  0.7   	&  -	    	&	-	 \\
Photon selection and kinematic fit    		&  -  		&  2.0 		&   1.0  	&	-	&	-    \\	
Extra photon energy requirement 			&  -  		&  0.6  	&   0.3   	&	-	&	-	 \\
Extra charged track requirement			&0.6 		&  0.6  	&    -    	&	-    	&	-	 \\
ST $M(\Ds)$ fit	         			  	&	0.9 	&  0.9  	&	 -	&	-	&	-	 \\
DT fit                                				& 0.8 		&   -   		&   0.4  	&	-	&	-	 \\
$MM^{2}$ fit                               			&  -  		&  1.5  	&   0.7  	&	-	&	-	 \\
MC statistics                         			& 0.3 		&  0.3  	&   0.2   	&	0.2 	&	 0.2 \\
Effect of $\br{}(\Dsst\ra\gam\Ds)		$		&  -		&  0.7  	&   0.3	  	&	-	&	-	 \\
Effect of $\ee\ra\Dsp\Dsm$        		& - 		&  0.4  	&	0.2	& -  	 	&  -    \\
Tag-side bias                              			& 0.3 		&  0.5  	&   0.3   	&	-	&	-	 \\
\hline
total                          		  		& 2.2 		&  3.1  	&  1.6  	& 0.6		&	0.6\\
\hline
\hline
\end{tabular}
\end{center}
\end{table*}

\subsection{Asymmetry measurement}

By using the measured branching fractions and  Eq.~(\ref{eq:RD}) the $\ksz$-$\klz$ asymmetry is determined to be 
\begin{eqnarray}
&R(\Dsp\ra \kz_{S,L} \kp) =  (-2.1~\pm~1.9_{\rm stat.})~\%. \label{eq:RDs_val}
\end{eqnarray}
To determine the direct $CP$ violation, we also measure the branching fractions for the $\Dsp$ and $\Dsm$ decays separately, using the same methodology as the combined branching fraction measurement. The direct $CP$ asymmetriy is defined as
\begin{equation}
A_{\rm CP}(\Dspm\ra f) = \frac{\br{}(\Dsp\ra f )-\br{}(\Dsm \ra \bar{f})}{\br{}(\Dsp\ra f )+\br{}(\Dsm\ra \bar{f})},
\label{eq:Acp}
\end{equation}
which leads to the measurements
\begin{eqnarray}
   A_{\rm CP}(\Dspm\ra\ksz\kpm) &=&   (\phantom{-}0.6~\pm~2.8_{\rm stat.})~ \%, \label{eq:AcpKsK}\\
   A_{\rm CP}(\Dspm\ra\klz\kpm)  &=&   (-1.1~\pm~2.6_{\rm stat.}) ~\%, \label{eq:AcpKLK}
\end{eqnarray}
for the two signal modes.

\section{SYSTEMATIC UNCERTAINTY}

 For the absolute branching fractions, which are determined according to Eq.~(\ref{eq:BF}), the systematic uncertainties are associated with $N_{\rm ST}^{i}$, $N_{\rm DT}^{\rm tot}$, and the corresponding ratio of detection efficiencies ($\epsilon_{\rm DT}^{i}/\epsilon_{\rm ST}^{i}$). One of the advantages of the DT method is that most of the systematic uncertainties associated with selection criteria for the ST side reconstruction cancel. However, there is some residual uncertainty due to the different decay topologies between DT and ST events; this is referred to as ``tag-side bias", and its effect is considered as one of the systematic uncertainties. For the $R(\Dsp)$ and $A_{\rm CP}$ measurements, the systematic uncertainties are calculated by propagating  corresponding branching fraction uncertainties from  different sources taking into account that some of the uncertainties cancel due to the fact that these observables are ratios as defined in Eqs.~(\ref{eq:RD}) and (\ref{eq:Acp}).

Table~\ref{tab:sys} summarizes the relative uncertainties on the absolute branching fraction and the absolute uncertainties for the asymmetries. The total systematic uncertainties are caculated as the sum in quadrature of individual contributions by assuming the sources are independent of one another.

The $\kp$ and $\km$ tracking efficiencies are studied using a control sample of $\ee\ra\kp\km\pip\pim$ events; the efficiency is calculated as a function of the transverse momentum of the particles. The average efficiency difference between data and MC is computed to be 0.5\% by weighting the efficiency difference found in the control sample according to the transverse momentum of kaon in signal MC samples. This is assigned as the systematic uncertainty from this source.

The $\kp$ and $\km$ PID efficiencies are studied using a control sample of $\Dsp\ra\kp\km\pip$, $\Dz\ra\km\pip$ and $\Dz\ra\km\pim\pip\pip$ events; the efficiency is calculated as a function of the momentum of the particle. The average efficiency difference between data and MC is computed to be 0.5\% by weighting the efficiency difference found in the control sample according to the momentum of kaon in signal MC samples, and this is assigned as the systematic uncertainty from this source. 

The $\ksz$ reconstruction efficiency has been studied using control samples of $J/\psi\ra K^{*}(892)^{\mp}K^{\pm}$ and $J/\psi\ra\phi\ksz\kpm\pimp$ in different momentum intervals~\cite{Ablikim:2015qgt}. The efficiency difference between data and MC is computed to be 1.5\%, which is assigned as the systematic uncertainty from this source.

 The systematic uncertainty associated with the photon selection efficiency and the kinematic fit in the study of $\Dsp\ra\klz\kp$ is estimated from the control sample $\Dsp\ra\pkkpi$. The same kinematic fit as that used on the data is performed by assuming the $\km\pip$ system is missing. The efficiency difference found between data and MC simulation, 2.0\%, is taken as the systematic uncertainty.
 
The systematic uncertainties associated with the requirements on the energy of additional photons and the number of extra charged tracks are estimated from the control sample $\Dsp\ra\pkkpi$. The efficiency differences between data and MC simulation for these two requirements are both 0.6\%, which are assigned as the systematic uncertainties from these sources. 

The uncertainty related to the limited sizes of MC samples is 0.3\%  for both $\Dsp\ra\pksk$ and $\Dsp\ra\klz\kp$.

The uncertainties associated with ST, DT, and $MM^{2}$ fits are studied by changing the signal and background PDFs, as well as the fit interval; each change is applied separately. Furthermore, in the $MM^{2}$ fit, the effect of the assumed peaking background yields is estimated by changing the fixed numbers of events by $\pm 1\sigma$. The systematic uncertainties related to the ST, DT, and $MM^{2}$ fit procedure are 0.9\%, 0.8\% and 1.5\%, respectively; these are the sums in quadrature of the relative changes of signal yield that result from each individual change to the fit procedure.
  
As discussed previously, the selected ST $\Dsm$ sample is dominated by the process $\ee\ra D_{s}^{*\pm}D_{s}^{\mp}\ra\gamma\Dsp\Dsm$, but there is small contribution from the processes $\ee\ra D_{s}^{*\pm}D_{s}^{\mp}\ra\piz\Dsp\Dsm$ and $\ee\ra\Dsp\Dsm$.
  In the analysis of $\Dsp\ra\ksz\kp$, detailed MC studies indicate that $\epsilon_{\rm DT}^i/\epsilon_{\rm ST}^i$ is almost the same for the three processes, since distributions of the kinematic variables are similar and no kinematic fit is performed in the DT selection. Thus, the effect from including $\ee\ra D_{s}^{*\pm}D_{s}^{\mp}\ra\piz\Dsp\Dsm$ and $\ee\ra\Dsp\Dsm$  processes is negligible in the absolute branching fraction measurement. In the analysis of $\Dsp\ra\klz\kp$, the kinematic fit is performed under the hypothesis that the event is $\ee\ra D_{s}^{*\pm}D_{s}^{\mp}\ra\gamma\Dsp\Dsm$, and the MC studies indicate that the contribution of $\ee\ra D_{s}^{*\pm}D_{s}^{\mp}\ra\piz\Dsp\Dsm$ and $\ee\ra\Dsp\Dsm$ in signal events can be neglected. Thus, the uncertainty of branching fraction $\br{}(\Dsstp\ra\gamma\Dsp)$~\cite{PDGweb} used in the signal MC simulation must be taken as a source of systematic uncertainty. The systematic uncertainty from excluding the process $\ee\ra\Dsp\Dsm$ is 0.4\%, which is the fraction of the ST yields that comes from the process $\ee\ra\Dsp\Dsm$; this fraction is estimated from the MC simulation.
  
The tag-side bias uncertainty is defined  as the uncanceled uncertainty in tag side due to different track multiplicities in generic and signal MC samples.  By studying the differences of tracking and PID efficiencies between data and MC in different multiplicities, the tag-side bias systematic uncertainties are estimated to be 0.3\% for $\Dsp\ra\pksk$ and 0.5\% for $\Dsp\ra\klz\kp$

\section{Summary and Discussion}

In summary, by using an $\ee$ collision data sample at $\sqrt{s}$ = 4.178~$\gev$, corresponding to an integrated luminosity of 3.19~$\ifb$, the absolute branching fractions are measured to be $\br{}(\Dsp\ra\pksk) = (1.425\pm0.038_{\rm stat.}\pm0.031_{\rm syst.})\%$ and $\br{}(\Dsp\ra\klz K^+) = (1.485\pm0.039_{\rm stat.}\pm0.046_{\rm syst.})\%$, the former  is one standard deviation lower than the world average value \cite{PDGweb}, and the latter is measured for the first time. The $\ksz$-$\klz$ asymmetry in $\Dsp$ decay  is measured for the first time as $R(\Dsp\ra \kz_{S,L}\kp)= (-2.1\pm1.9_{\rm stat.}\pm1.6_{\rm syst.})\%$. This mearsurement is compatible with theoretical predictions listed in Table~\ref{tab:theorypre}.  Direct $CP$ asymmetries of the two processes are obtained to be $A_{\rm CP}(\Dspm\ra\ksz\kpm) =  (0.6\pm2.8_{\rm stat.}\pm0.6_{\rm syst.})\%$ and $A_{\rm CP}(\Dspm\ra\klz\kpm) = (-1.1\pm2.6_{\rm stat.}\pm0.6_{\rm syst.})\%$. No significant asymmetries are observed and the uncertainties are statistically dominant.

\section{acknowledgments}

The BESIII collaboration thanks the staff of BEPCII and the IHEP computing center and the supercomputing center of USTC  for their strong support. This work is supported in part by National Key Basic Research Program of China under Contract No. 2015CB856700; National Natural Science Foundation of China (NSFC) under Contracts Nos. 11335008, 11375170, 11475164, 11475169, 11605196, 11605198, 11625523, 11635010, 11705192, 11735014; National Natural Science Foundation of China (NSFC) under Contract No. 11835012; the Chinese Academy of Sciences (CAS) Large-Scale Scientific Facility Program; Joint Large-Scale Scientific Facility Funds of the NSFC and CAS under Contracts Nos. U1532102, U1532257, U1532258, U1632109, U1732263, U1832207; CAS Key Research Program of Frontier Sciences under Contracts Nos. QYZDJ-SSW-SLH003, QYZDJ-SSW-SLH040; 100 Talents Program of CAS; INPAC and Shanghai Key Laboratory for Particle Physics and Cosmology; German Research Foundation DFG under Contract No. Collaborative Research Center CRC 1044; Istituto Nazionale di Fisica Nucleare, Italy; Koninklijke Nederlandse Akademie van Wetenschappen (KNAW) under Contract No. 530-4CDP03; Ministry of Development of Turkey under Contract No. DPT2006K-120470; National Science and Technology fund; The Knut and Alice Wallenberg Foundation (Sweden) under Contract No. 2016.0157; The Swedish Research Council; U. S. Department of Energy under Contracts Nos. DE-FG02-05ER41374, DE-SC-0010118, DE-SC-0012069; University of Groningen (RuG) and the Helmholtzzentrum fuer Schwerionenforschung GmbH (GSI), Darmstadt; WCU Program of National Research Foundation of Korea under Contract No. R32-2008-000-10155-0.



\end{document}